\documentclass{article}
\usepackage{amssymb}
\usepackage{amsbsy}
\usepackage{epsf}
\usepackage{epsfig}
\usepackage{amsfonts}
\usepackage{latexsym}
\usepackage[mathscr]{eucal}
\usepackage{amscd}
\usepackage[centertags]{amsmath}
\usepackage{dsfont}
\usepackage{yfonts} 
\usepackage{enumerate}
\usepackage[center]{subfigure}
\usepackage{cite}

\newcommand{\beq}{\begin{equation}}
\newcommand{\eeq}{\end{equation}}
\newcommand{\beqn}{\begin{eqnarray}}
\newcommand{\eeqn}{\end{eqnarray}}
\newcommand{\ol}{\overline}
\newcommand{\wt}{\widetilde}
\newcommand{\bs}{\boldsymbol}
\newcommand{\m}{\mathcal}

\begin{document}

\title{\hfill ~\\[-30mm]
       \hfill\mbox{\small UFIFT-HEP-08-10}\\[30mm]
       \textbf{Discrete Anomalies of Binary Groups}}

\date{}
\author{\\Christoph Luhn\footnote{E-mail: {\tt luhn@phys.ufl.edu}}\\[3mm]
  \emph{\small{}Institute for Fundamental Theory, Department of Physics,}\\
  \emph{\small University of Florida, Gainesville, FL 32611, USA}}

\maketitle

\begin{abstract}
\noindent We derive the discrete anomaly conditions for the binary tetrahedral
group~$\m T'$ as well as the binary dihedral groups~$\m Q_{2n}$. The ambiguities
of embedding these finite groups into $SU(2)$ and $SU(3)$ lead to various
possible definitions of the discrete indices which enter the anomaly
equations. We scrutinize the different choices and show that it is sufficient to
consider one particular assignment for the discrete indices. Thus it is
straightforward to determine whether or not a given model of flavor is discrete
anomaly free. 
\end{abstract}

\thispagestyle{empty}
\vfill
\newpage
\setcounter{page}{1}


\section{Introduction}

The family structure of the Yukawa couplings which give rise to the masses and
mixings of quarks and leptons remains unexplained within the Standard
Model (SM). One of the most successful ideas to overcome this annoying
shortcoming consists in extending the SM gauge group with a family-dependent
$U(1)$ symmetry. These Froggatt-Nielsen~\cite{Froggatt:1978nt} models seem to
be well suited for addressing the hierarchies of the quark and charged lepton
masses as well as the small angles of the Cabibbo-Kobayashi-Maskawa (CKM) mixing
matrix. However, they fail to give a convincing account of the neutrino sector
which features either no or only a very mild hierarchy. 

The observation that the Maki-Nakagawa-Sakata-Pontecorvo (MNSP) mixing matrix,
to a very good approximation, exhibits the so-called tri-bimaximal
structure~\cite{Harrison:2002er} has
added its share to the mystery surrounding the fermionic masses and
mixings. Spurred on by this remarkable fact, model builders have resorted to
imposing an underlying non-Abelian finite family symmetry. As there are three
families of quarks and leptons, the finite group should have two- or
three-dimensional irreps. This requirement limits the candidates to the finite
subgroups of $SU(3)$, $SO(3)$, and $SU(2)$.\footnote{An embedding of these
finite groups into bigger continuous groups is possible in principle, but it
would not be a genuine one.} 
Having picked a preferred finite 
group $\m G$, one is still left with the choice of assigning the various
particles of a model to irreps of $\m G$. 

In order to constrain the possibilities,
Ref.~\cite{Luhn:2008sa} exploits the conditions arising from requiring that
$\m G$ should originate in an anomaly free gauge symmetry, a framework which
stabilizes the discrete symmetry against violation by quantum gravity
effects~\cite{Krauss:1988zc}. The  formulation of the discrete anomaly
conditions necessitates the 
definition of Dynkin-type indices for the irreps of finite groups. These
so-called discrete indices have to be defined individually for each group~$\m
G$.  Focusing on several popular finite family groups, the general
procedure of obtaining the discrete anomaly conditions has been established in
Ref.~\cite{Luhn:2008sa}. In all cases, $\m G$ was 
a subgroup of $SU(3)$ or $SO(3)$ but not $SU(2)$. Furthermore, the embedding
into the continuous group was always defined uniquely.\footnote{In the case of
  $\m D_5$ there are actually two distinct embeddings. However, they are
  equivalent as they can be related to each other by relabeling the
  two-dimensional irreps ${\bf 2_1 \leftrightarrow 2_2}$.} 

It is the purpose of this letter to discuss the constraint arising from the
discrete anomaly $\m G-\m G-U(1)_Y$ for some finite subgroups of $SU(2)$ which
have been put forward as possible 
family symmetries, notably the binary tetrahedral group $\m
T'$~\cite{Aranda:2000tm,Chen:2007afa,Feruglio:2007uu,Frampton:2007et,Ding:2008rj,Frampton:2007hs}
and the binary dihedral groups $\m
Q_{2n}$~\cite{Chang:1990uga,Chang:1991ri,Frampton:1994xm,Frampton:1994rk,
Frampton:1995fta,Frampton:1995gu,Frampton:1996cc,
Babu:2004tn,Frigerio:2004jg,Frigerio:2005pz,Kajiyama:2005rk}. The fact that subgroups of $SU(2)$
necessarily have two-dimensional irreps appears to have some advantage since a
$2+1$ structure can naturally separate the third family of fermions from the
other two. 
In the process of deriving the discrete anomaly conditions we will find that
there exist different embeddings which seem to make a consistent definition of
the discrete indices impossible. It is one of the main intentions of this
letter to shed some light on this ambiguity, proving that {\it one particular
assignment of discrete indices is sufficient} to determine whether a specific
model of flavor is discrete anomaly free or not. We illustrate the procedure
by applying our results to some existing examples.

\section{The Binary Tetrahedral Group $\bs{\m T'}$}

The alternating group on four letters, $\m A_4$, is the symmetry group of the
tetrahedron, and as such a subgroup of~$SO(3)$. It has three one-dimensional
and one three-dimensional irreps. Similar to~$SU(2)$
being the double cover of~$SO(3)$, the binary tetrahedral group~$\m T'$ is
the double cover of~$\m A_4$. It is a subgroup of~$SU(2)$ and has three
two-dimensional irreps in addition to those of~$\m A_4$. A
convenient way to define this group is provided in terms of its
presentation~\cite{Thomas,Feruglio:2007uu}
\beq
\langle r,s,t \,|\, r^2 = t^3 = (st)^3 = 1,  \, s^2=r, \, rt=tr \rangle \
.\label{T'pres} 
\eeq
Here we have introduced the auxiliary generator $r$ in order to manifest the
connection between the binary tetrahedral group and the alternating group:
setting $r=1$ in Eq.~(\ref{T'pres}) yields the presentation of~$\m A_4$. The
one- and three-dimensional irreps of~$\m T'$ are
identical to the irreps of~$\m A_4$
$$
\begin{array}{rl}
{\bf 1_k}: ~~& s~=~1 \ , \qquad  t~=~\omega^k \ , \\[4mm] 
{\bf 3~}: ~~ & s~=\,\begin{pmatrix} 1  & 0&0 \\ 0&-1 &0 \\0&0&-1 \end{pmatrix}  \ ,
\qquad  t~=\,\begin{pmatrix} 0& 1& 0 \\ 0&0 & 1 \\ 1 &0 &0  \end{pmatrix} \ , 
\end{array}
$$
while the two-dimensional irreps take the form 
$$
\begin{array}{rll}
{\bf 2_k}: ~~ & s~=~\begin{pmatrix} i  & 0 \\ 0 & -i\end{pmatrix}  \ , \qquad &
t~=~ \frac{\omega^k}{\sqrt{2}} \begin{pmatrix} \tau^3 & \tau \\ \tau^3 &
  \tau^5  \end{pmatrix} \ , ~~
\end{array}
$$
with $\omega =e^{2 \pi i/3}$, $\tau=e^{\pi i/4}$ and  $k=0,1,2$. The
resulting Kronecker products are those of~$\m A_4$
$$
{\bf 1_k} \otimes {\bf 1_l} ~=~ {\bf 1_{k+l}} \ , \qquad 
{\bf 1_k} \otimes {\bf 3} ~=~ {\bf 3} \ , \qquad
{\bf 3} \otimes {\bf 3} ~=~ {\bf 1_0} + {\bf 1_1}+{\bf 1_2} + 2 \cdot {\bf 3}
\ ,
$$
plus the ones involving the doublets
$$
{\bf 2_k} \otimes {\bf 1_l} ~=~ {\bf 2_{k+l}} \ , \qquad 
{\bf 2_k} \otimes {\bf 2_l} ~=~ {\bf 1_{k+l}}+{\bf 3} \ , \qquad
{\bf 2_k} \otimes {\bf 3} ~=~ {\bf 2_0} + {\bf 2_1}+{\bf 2_2} \ .
$$
The subscripts are understood modulo~3, so that the complex conjugates of
${\bf 1_k}$ and ${\bf 2_k}$ are given as ${\bf \ol 1_k}={\bf 1_{-k}}$ and
${\bf \ol 2_k}={\bf 2_{-k}}$. Writing the spinor of the irrep ${\bf 2_k}$ as
$\begin{pmatrix} u \\  v \end{pmatrix}$, the conjugated spinor 
$\begin{pmatrix}v^\ast \\ -u^\ast\end{pmatrix} = i \sigma_2 \begin{pmatrix}
  u^\ast \\  v^\ast \end{pmatrix}$ transforms with the same matrices as the
irrep ${\bf 2_{-k}}$. Notice that for $k=0$ this is similar to the ${\bf 2}$
of $SU(2)$ being its own conjugate.

Since the determinant\footnote{Bear in mind that the generators $r$, $s$, $t$
  of the finite group shall be elements of the $SU(2)$ Lie {\it group} and not
  of the $SU(2)$ Lie {\it algebra}.}  
of $t$ is $\omega^{2k}$ for the two-dimensional irreps,
there is only one embedding of $\m T'$ into $SU(2)$: the 
${\bf 2}$ of $SU(2)$ is identified with the~${\bf 2_0}$~of~$\m{T}'$. Then the
decomposition of all other irreps of $SU(2)$ is fixed by the Kronecker
products. The spinors break up into sums of ${\bf 2_k}$, while the vectors
decompose into sums of ${\bf 1_k}$ and ${\bf 3}$. Hence the two-dimensional
irreps of $\m T'$ are spinor-like. The following table lists the decomposition
of the smallest irreps ${\bs \rho}$ of $SU(2)$ into irreps ${\bf r}$ of $\m
T'$. For the purpose of assigning discrete indices to ${\bf r}$, we
also display the Dynkin indices $\ell({\bs \rho})$.
\begin{center}
\begin{tabular}{c|c|c}
&& \\[-3mm]
\begin{tabular}{c} Irreps ${\bs \rho}$ \\ of $SU(2)$ \end{tabular}
&\begin{tabular}{c} Decomposition\\ of ${\bs \rho}$ under $\m T'$\end{tabular} 
&\begin{tabular}{c} Dynkin \\ index $\ell({\bs \rho})$ \end{tabular}  \\[-3mm] 
& &\\ \hline && \\[-3mm]
{\bf 2}  & ${\bf 2_0}$ & 1   \\
{\bf 3}  & ${\bf 3~}$ & 4  \\
{\bf 4}  & ${\bf 2_1}+{\bf 2_2}$ & 10   \\
{\bf 5}  & ${\bf 1_1}+{\bf 1_2}+{\bf 3}$ & 20  \\
{\bf 6}  & ${\bf 2_0}+{\bf 2_1}+{\bf 2_2}$ & 35  \\
{\bf 7}  &${\bf 1_0}+2 \cdot {\bf 3}$& 56    \\[1mm]
\end{tabular}
\end{center}

\vspace{1mm}

\noindent It is easy to show that ${\bf 1_1}$ and ${\bf 1_2}$ as well as
 ${\bf 2_1}$ and ${\bf 2_2}$ always come in pairs, so that the discrete
indices $\wt \ell({\bf r})$ cannot be defined uniquely. We obtain
\beq
\begin{array}{llll}
\wt \ell({\bf 1_0}) ~=~0\ , &\quad \wt \ell({\bf 1_1}) ~=~x\ , &\quad \wt
\ell({\bf 1_2}) ~=~16-x\ , & \\[2mm]
\wt \ell({\bf 2_0}) ~=~1\ , &\quad \wt \ell({\bf 2_1}) ~=~y\ , &\quad \wt
\ell({\bf 2_2}) ~=~10-y\ , & \quad \wt \ell({\bf 3}) ~=~4\ ,\\
\end{array} \label{T'indices}
\eeq
where the parameters $x$ and $y$ can take arbitrary values.  
Comparing the decomposition of the ${\bf 6}$ with those of the ${\bf 2}$ and
${\bf 4}$, we see that the discrete indices $\wt \ell({\bf r})$ can only be
defined modulo~$N_\ell=24$. Using the methods presented in
Ref.~\cite{Luhn:2008sa} one can prove that the assignments in
Eq.~(\ref{T'indices}) are consistent for all irreps of~$SU(2)$.
That is, given an arbitrary $SU(2)$ irrep ${\bs{\rho}}$ and its decomposition into
irreps of the finite subgroup, the sum of the corresponding {\it discrete}
indices adds up to the Dynkin index of ${\bs{\rho}}$ {\it modulo~$N_\ell$}, see also
Eq.~(2.6) of Ref.~\cite{Luhn:2008sa}. 

At this point, due to the
modulo~$N_\ell$, it is already evident that the discrete anomaly conditions
derived in the following are {\it necessary but not sufficient} for
a given model of flavor to originate from an anomaly free gauged flavor
symmetry. In order to guarantee that the continuous high-energy flavor theory
is anomaly free, the full theory together with its various breaking 
mechanisms and the resulting heavy degrees of freedom need to be investigated
in detail. Such an endeavor goes well beyond the scope of this letter. 

It is worth mentioning that there is another condition (independent from the
anomaly discussion) which is necessary in order to gauge the discrete
symmetry: The continuous high-energy  
theory must always, by definition, start with {\it complete} irreps of
$SU(2)$. This, however, does not  entail that the {\it light} irreps of $\m G$
have to add up to complete $SU(2)$ irreps, because parts of an original
$SU(2)$ irrep might acquire a mass while the rest remains massless. For
instance, consider the ${\bf 5}$ of $SU(2)$ which decomposes into ${\bf 1_1 +
  1_2 + 3}$ of~$\m T'$. The one-dimensional irreps might remain light, while
the triplet acquires a mass. This could be achieved by introducing a new ${\bf
  3}$ of $SU(2)$  which, after the breakdown to $\m T'$, transforms as a ${\bf
  3}$ of $\m T'$ and can then form a bilinear mass term with the triplet of
the original~${\bf   5}$. This situation is in some sense analogous to the
doublet-triplet splitting in $SU(5)$ grand unified theories.

Nonetheless, there do exist cases in which it is possible to show that the
assignment of the light particle content under the discrete symmetry {\it
  cannot} originate from complete multiplets of $SU(2)$ unless additional
light degrees of freedom are introduced. This is owed to the
breaking pattern of $SU(2)$ down to~$\m T'$: the irreps ${\bf 1_1}$ and ${\bf
  1_2}$ {\it always} come in pairs, as do the irreps ${\bf 2_1}$ and ${\bf
  2_2}$. Since mass terms need to be of the form ${\bf 1_1} \otimes {\bf 1_2}$
or  ${\bf 2_1} \otimes {\bf 2_2}$, it is impossible to make one constituent of
such a pair of irreps heavy while the other remains light (without having a
new light field which again would complete the pair). Therefore, the models of
Refs.~\cite{Feruglio:2007uu,Frampton:2007et,Frampton:2007hs,Ding:2008rj} are
incomplete within an $SU(2)$ framework\footnote{In the case where $\m T'$
  originates from $SU(3)$, the situation is more involved due to different
  allowed breaking patterns of the $SU(3)$ irreps. As it is the intention of
  this letter to 
  discuss the discrete anomaly conditions, we just state that 
  all models of Table~\ref{tabT} are incomplete because of the requirement to
  start out with complete multiplets of either $SU(2)$ or $SU(3)$. Yet, we 
  show the examples to illustrate the procedure of applying our discrete anomaly
  conditions.} already from this perspective.

Turning back to the discussion of the anomalies, one can alternatively embed
the binary tetrahedral group into $SU(3)$ instead of $SU(2)$. Identifying the
${\bf 3}$ of $SU(3)$ with the ${\bf 3}$ of $\m T'$ 
would only generate the representations of $\m A_4$, excluding the
two-dimensional spinor-like irreps of $\m T'$. We are therefore left with
three conceivable embeddings which are defined by:
$$
(i):~{\bf 3}~\rightarrow~{\bf 1_0}+{\bf 2_0} \ , \qquad
(ii):~{\bf 3}~\rightarrow~{\bf 1_1}+{\bf 2_1} \ , \qquad
(iii):~{\bf 3}~\rightarrow~{\bf 1_2}+{\bf 2_2} \ .
$$
As in the case of $SU(2)$, the derived discrete indices are not uniquely
determined. For $(i)$, the assignments turn out to be identical to those of
Eq.~(\ref{T'indices}). On the other hand, $(ii)$ and $(iii)$ lead to the
discrete indices of Eq.~(\ref{T'indices}) with $y$ replaced by $1-x$.

The question arises: Which  indices should be used for the discrete
anomaly conditions? Since the low-energy models that adopt a discrete family
symmetry do not specify a particular embedding, the most general approach
consists in choosing a parameterization of the discrete indices which holds
for {\it any} of the above embeddings. That is, we have to set $y=1-x$ in
Eq.~(\ref{T'indices}). Consequently, the discrete indices solely depend on the
parameter~$x$.

The requirement of massive degrees of freedom not affecting the discrete
anomaly conditions further constrains~$x$. 
Among the particles that have a $\m T'$~invariant mass term, only the pairs
${\bf 1_1}$ and ${\bf 1_2}$ 
as well as ${\bf 2_1}$ and ${\bf 2_2}$ might give non-zero contributions to the
$\m T'-\m T'-U(1)_Y$ anomaly. With the hypercharges being opposite to each
other, the particles of each pair add
$$
\begin{array}{ll}
{\bf 1_1}-{\bf 1_2}: & \quad Y_{\bf 1_1}  (2x-16) \ ,\\[2mm]
{\bf 2_1}-{\bf 2_2}: & \quad Y_{\bf 2_1}  (2y-10) \,=\, - Y_{\bf 2_1} 
(2x+8) \ ,
\end{array}
$$
to the discrete anomaly. Choosing either $x=8$ or $x=20$ their
contributions vanish modulo~24 provided that the hypercharges $Y_{...}$ are
normalized to be integer.\footnote{See Footnote~5 of Ref.~\cite{Luhn:2008sa}
  for a discussion of the convention concerning the hypercharge normalization.}
Therefore a model which features a $\m T'$ family symmetry is discrete anomaly
free and thus consistent with the assumed light particle content if 

\vspace{-1.0mm}

\beq
\sum_{i=\mathrm{light}} Y_i \cdot \wt \ell_i  ~=~ 0~\mathrm{mod}~24\ ,\label{Tdac}
\eeq

\vspace{-1.0mm}

\noindent is satisfied for the discrete indices shown in Table~\ref{tabDIT}, with
$\xi=0,1$. We emphasize that the discrete anomaly condition of
Eq.~(\ref{Tdac}) is independent of the possible embeddings into $SU(2)$ or
$SU(3)$. 
\begin{table}[t]
\begin{center}
\begin{tabular}{|c|c|}
\hline  & \\[-3mm]
\begin{tabular}{c} $\m T'$ \\ irreps \end{tabular}
& \begin{tabular}{c} $\wt \ell({\bf r})$ \\ $[\,N_\ell = 24$\,]\end{tabular}
\\[-3mm] 
& \\ \hline & \\[-3mm]
${\bf 1_0}$  & 0     \\
${\bf 1_1}$  & $8 \,+\,\xi\cdot 12$      \\
${\bf 1_2}$  & $8\,+\,\xi\cdot 12$      \\
${\bf 2_0}$  & 1     \\
${\bf 2_1}$  & $17\,-\,\xi\cdot 12$     \\
${\bf 2_2}$  & $17\,-\,\xi\cdot 12$     \\
${\bf 3} $   & 4    \\[1mm]\hline
\end{tabular}
\end{center}\vspace{-2.5mm}
\caption{\label{tabDIT}The discrete indices $\wt \ell$ for the irreps ${\bf
    r}$ of $\m T'$, with $\xi =0,1$.}\vspace{-1.5mm}
\end{table}

Let us now apply Eq.~(\ref{Tdac}) to some existing models of
flavor. Demanding a grand unified structure in which all particles of the
$SU(5)$ multiplets transform identically under $\m T'$, the models of
Refs.~\cite{Aranda:2000tm,Chen:2007afa} are automatically discrete
anomaly free because the sum of the hypercharges vanishes within each
multiplet. In
Refs.~\cite{Feruglio:2007uu,Frampton:2007et,Frampton:2007hs,Ding:2008rj} the 
only particles contributing to the discrete anomaly of Eq.~(\ref{Tdac}) are
the quarks and leptons. Their assignments to irreps of $\m T'$ are given 
in Table~\ref{tabT}, showing that the models of
Refs.~\cite{Feruglio:2007uu,Frampton:2007et,Ding:2008rj}  
are discrete anomaly free, while
the one in Ref.~\cite{Frampton:2007hs} is anomalous and therefore
incomplete.\footnote{Shortly after submitting this article to the archive,
  Ref.~\cite{Frampton:2007hs} was withdrawn by the authors.}
\begin{table}[h]
\begin{center}
\begin{tabular}{|c|c|c|c|c|c|c|}
\hline  & &&&&& \\[-3mm]
$\m T'$ models & $Q$ & $u^c$ &  $d^c$ & $L$ & $e^c$  & $\sum Y_i \, \wt \ell_i$ \\[-3mm] 
&&&&&&\\ \hline &&&&&& \\[-3mm]
Ref.~\cite{Feruglio:2007uu} & ${\bf 1_0 , 2_2}$ & ${\bf 1_0 , 2_2}$ & ${\bf
  1_0 , 2_2}$ & ${\bf 3}$ & ${\bf 1_0 , 1_1 , 1_2}$ & $0~\mathrm{mod}~24$
\\[-3mm] 
&&&&&&\\ \hline &&&&&& \\[-3mm] 
Ref.~\cite{Frampton:2007et} & ${\bf 1_0 , 2_0}$ & ${\bf 1_0 , 2_1}$ & ${\bf
  1_1 , 2_2}$ & ${\bf 3}$ & ${\bf 1_0 , 1_1 , 1_2}$ & $0~\mathrm{mod}~24$
\\[-3mm] 
&&&&&&\\ \hline &&&&&& \\[-3mm] 
Ref.~\cite{Ding:2008rj} & ${\bf 1_2 , 2_1}$ & ${\bf 1_1 , 2_0}$ & ${\bf
  1_1 , 2_0}$ & ${\bf 3}$ & ${\bf 1_0 , 1_1 , 1_2}$ & $0~\mathrm{mod}~24$
\\[-3mm] 
&&&&&&\\ \hline &&&&&& \\[-3mm] 
Ref.~\cite{Frampton:2007hs} & ${\bf 1_0 , 2_0}$ & ${\bf 1_0 , 1_1 , 1_2}$ &
${\bf 1_1 , 2_2}$ & ${\bf 3}$ & ${\bf 1_0 , 1_1 , 1_2}$ & $12~\mathrm{mod}~24$ \\[1mm]\hline
\end{tabular}
\end{center}\vspace{-2.5mm}
\caption{\label{tabT}The $\m T'-\m T'-U(1)_Y$ anomaly for various flavor
  models adopting the shown assignments of the quarks and leptons to irreps of
  $\m T'$.} \vspace{2mm}
\end{table}

Before turning to the binary dihedral groups $\m Q_{2n}$, we should
briefly remark on possible remnants of the Witten $SU(2)$
anomaly~\cite{Witten:1982fp}. Mathematical consistency of the theory requires
the number of left-handed fermionic $SU(2)$ doublets to be even if higher
$SU(2)$ irreps are absent. Allowing for arbitrary $SU(2)$ irreps, this
statement generalizes to the condition 
that the number of left-handed fermionic irreps with odd Dynkin index must be
even, see e.g. Ref.~\cite{Geng:1987fg}. These are the $SU(2)$ irreps of
dimension $2+4m$ with $m\in \mathbb N$. As these $SU(2)$ irreps are the only
ones that, under $\m T'$, decompose with an odd number of ${\bf 2_0}$ irreps,
the constraint from the Witten anomaly breaks down to  
the requirement of having an even number of ${\bf 2_0}$ irreps in the complete
theory, including the heavy degrees of freedom. From the Kronecker
products we see that a heavy Majorana particle transforming as a ${\bf 2_0}$
of $\m T'$ is possible. Therefore the constraint from the Witten anomaly does
not leave its footprints on the light particle content of a $\m T'$ symmetric
theory.


\section{The Binary Dihedral Groups $\bs{\m Q_{2n}}$}

The dihedral group $\m D_n$ is the symmetry group of the planar $n$-polygon.
Its $2n$ elements can be expressed as rotations in three-dimensional space,
indicating that $\m D_n \subset SO(3)$. The dicyclic or binary dihedral group
$\m Q_{2n}$ is the double cover of $\m D_n$. It is defined by the
presentation~\cite{Thomas,Blum:2007jz}
\beq
\langle r,a,b \,|\, r^2 = 1,  \, a^n = b^2 = r, \, a b a =b \rangle \
.\label{Qpres} 
\eeq
Setting the auxiliary generator $r$ to one, we recover the presentation of $\m
D_n$. The irreps of the binary dihedral group $\m Q_{2n}$ are
$$
\begin{array}{rl}
{\bf 1_{0,1}}: ~~& a~=~1 \ , ~~~\:\qquad  b~=~\pm 1 \ , \\[4mm] 
{\bf 1_{2,3}}: ~~& a~=~-1 \ , \qquad  b~=\,\left\{ 
\begin{array}{ll} \pm 1  & ~~ (n=\mathrm{even})\ , \\[2mm]  \pm i  &~~
  (n=\mathrm{odd}) \ ,   \end{array} \right.  \\[7mm]
{\bf 2_k}: ~~ & a~=\,\begin{pmatrix} \eta^k & 0 \\ 0&\eta^{-k} \end{pmatrix}
\ , \qquad  b~=\,\begin{pmatrix} 0& 1 \\  (-1)^k &0  \end{pmatrix} \ , 
\end{array}
$$
where $\eta=e^{\pi i/ n}$. The parameter $k$ labels the two-dimensional
representations and can formally take any integer value. However, a closer
look at the generators reveals that  
\beq
{\bf 2_{-k}} ~=~ {\bf 2_{k}} ~=~ {\bf 2_{k+2n} } \ ,\label{ident1}
\eeq
denote identical representations. Furthermore, $k=0$ and $k=n$ lead to
reducible representations:
\beq
{\bf 2_0} ~=~ {\bf 1_0 \,+\, 1_1} \ , \qquad {\bf 2_n} ~=~ {\bf 1_2 \,+\, 1_3} \ .\label{ident2}
\eeq
Therefore there are $n-1$ inequivalent irreducible two-dimensional
representations, labeled by $k=1,...,n-1$. The irreps of $\m D_n$ can be
easily identified as those ${\bf 2_k}$ with even $k$, including ${\bf 2_0}$
and (if $n$ is even) ${\bf 2_n}$. This shows that, for odd $n$, $\m D_n$ has
only two one-dimensional irreps.

The Kronecker products for the irreps of $\m Q_{2n}$ are
$$
{\bf 1_1} \otimes {\bf 1_1} ~=~ {\bf 1_{0}} \ , \qquad 
{\bf 1_1} \otimes {\bf 1_2} ~=~ {\bf 1_3} \ , \qquad
{\bf 1_1} \otimes {\bf 1_3} ~=~ {\bf 1_2} \ ,\\[3mm]
$$
$$
{\bf 1_{2}} \otimes {\bf 1_{2}}~=~{\bf 1_{3}} \otimes {\bf 1_{3}} ~=\,\left\{ \begin{array}{ll} {\bf 1_0} \  &  (n=\mathrm{even}) \ ,\\[2mm] {\bf 1_1}  & (n=\mathrm{odd}) \ ,  \end{array} \right. 
\quad
{\bf 1_{2}} \otimes {\bf 1_{3}} ~=\,\left\{ \begin{array}{ll} {\bf 1_1} \  &  (n=\mathrm{even}) \ ,\\[2mm] {\bf 1_0}  & (n=\mathrm{odd}) \ ,  \end{array} \right. \\[4mm]
$$
$$
{\bf 1_1} \otimes {\bf 2_k} \,=\, {\bf 2_{k}} \ , ~~\quad 
{\bf 1_2} \otimes {\bf 2_k} \,=\,{\bf 1_3} \otimes {\bf 2_k} \,=\, {\bf 2_{n-k}} \ , ~~\quad{\bf 2_k} \otimes {\bf 2_l} \,=\,{\bf 2_{k+l}} + {\bf 2_{k-l}} \ ,
$$
where we make use of the identities in Eqs.~(\ref{ident1}) and~(\ref{ident2}).
It is worth mentioning that the products involving ${\bf 2_k}$ remain valid
for $k=0$ or $n$.

We now wish to embed $\m Q_{2n}$ into $SU(2)$. Since the determinant of the
generator $b$ for two-dimensional irreps depends on $k$, only ${\bf 2_k}$
with an odd value for $k$ can be identified with the ${\bf 2}$ of
$SU(2)$. They are the spinor-like irreps of $\m Q_{2n}$, whereas the ${\bf 2_k}$
with even $k$ are vector-like. Furthermore, we want to embed the whole group
$\m Q_{2n}$ and not just a subgroup of it. For that reason, the ${\bf 2}$ (which
generates all other $SU(2)$ irreps by successive multiplication) must
correspond to ${\bf 2_{\bs \alpha}}$ with $\alpha$ and $2n$ being coprime,
i.e. they have no common prime factor. For example, with $n=10$ there are
four possibilities, namely $\alpha=1,3,7,9$.
Having identified the ${\bf 2}$ of $SU(2)$ with ${\bf 2_{\bs\alpha}}$  of $\m
Q_{2n}$, the decomposition of the other irreps of $SU(2)$ is obtained from the
Kronecker products. Using Eqs.~(\ref{ident1}) and (\ref{ident2}), one
finds the following embedding~\cite{Frampton:1999hk}.\footnote{The constraints
  arising from the Witten anomaly require that the number of spinor-like irreps,
  i.e. the ${\bf {2_{\kappa\alpha}}}$ with $\kappa=\mathrm{odd}$, is even if all
  degrees of freedom are counted. However, the Kronecker products show that
  these spinor-like $\m Q_{2n}$ irreps can form Majorana mass term. Therefore,
  as in the case of $\m T'$, no constraint on the light particle content is
  obtained from the Witten anomaly.} 
\begin{center}
\begin{tabular}{c|c|c}
&& \\[-3mm]
\begin{tabular}{c} Irreps ${\bs \rho}$ \\ of $SU(2)$ \end{tabular}
&\begin{tabular}{c} Decomposition\\ of ${\bs \rho}$ under $\m Q_{2n}$\end{tabular} 
&\begin{tabular}{c} Dynkin \\ index $\ell({\bs \rho})$ \end{tabular}  \\[-3mm] 
& &\\ \hline && \\[-3mm]
{\bf 2}  & ${\bf 2_{\bs\alpha}}$ & 1   \\
{\bf 3}  & ${\bf 1_1 + 2_{2\bs\alpha}}$ & 4  \\
{\bf 4}  & ${\bf 2_{\bs\alpha}}+{\bf 2_{3\bs\alpha}}$ & 10   \\
{\bf 5}  & ${\bf 1_0}+ {\bf 2_{2\bs\alpha}}+{\bf 2_{4\bs\alpha}}$ & 20  \\
{\bf 6}  & ${\bf 2_{\bs\alpha}}+{\bf 2_{3\bs\alpha}}+{\bf 2_{5\bs\alpha}}$ & 35  \\
{\bf 7}  &${\bf 1_1}+{\bf 2_{2\bs\alpha}}+{\bf 2_{4\bs\alpha}}+{\bf 2_{6\bs\alpha}} $& 56    \\[1mm]
\end{tabular}
\end{center}

\vspace{1mm}

\noindent Again the Dynkin indices of $SU(2)$ are shown in order to help
extract the discrete indices~$\wt \ell({\bf r})$ of the irreps of $\m
Q_{2n}$. Setting $\wt \ell({\bf 2_0})=\wt \ell({\bf 1_1})=x$, it is an easy
matter to prove that, in a compact notation,
\beq
\wt \ell({\bf 2_{\bs{\kappa\alpha}}}) ~=~  \kappa^2 \,+\, \frac{x}{2}
\cdot(i^\kappa + i^{-\kappa})  \ , \label{DIQ}
\eeq
is consistent for the decomposition of all $SU(2)$ irreps. Notice that for odd
$\kappa$ the $x$-dependence drops out, while even $\kappa$ entails the
term $x i^\kappa$ in addition to $\kappa^2$. Of course, the discrete indices are
only defined modulo~$N_\ell$. In oder to determine its value we have to
recall that ${\bf 2_{(n+l)\bs\alpha} = 2_{(n-l)\bs\alpha}}$ so that their discrete
indices must be identical, i.e. 
$$
\wt \ell({\bf 2_{(n+l)\bs\alpha}}) \,-\, \wt \ell({\bf 2_{(n-l)\bs\alpha}})
~=~4n l + \frac{x}{2} \cdot(i^{n+l}+i^{-n-l} - i^{n-l} - i^{-n+l}) ~=~
0~\mathrm{mod}~N_\ell \ ,
$$
for all $l=1,...,n$. This results in $N_\ell = 4 n$.  While the sum in the
parentheses vanishes for even $n$, it is non-vanishing for odd $n$. In the
latter case, $x$ is therefore additionally constrained to be either $0$ or
$2n$, so that the discrete indices for 
the irreps of the binary dihedral group $\m Q_{2n}$ with odd $n$ are given by 
\beq
\wt \ell({\bf 2_{\bs{\kappa\alpha}}}) ~=~  \kappa^2 \,+\, \xi \cdot n\,
(i^\kappa + i^{-\kappa})  
\qquad(n=\mathrm{odd}) \ ,\label{DIQodd}
\eeq
with $\xi=0,1$. For even $n$, Eq.~(\ref{DIQ}) remains unchanged. 
Notice that the discrete indices for ${\bf 2_{\bs{\kappa\alpha}}}$,
${\bf 2_{\bs{-\kappa\alpha}}}$, and ${\bf 2_{\bs{(\kappa+2n)\alpha}}}$ are
identical as required by the identities of Eq.~(\ref{ident1}). 
We emphasize that different embeddings of $\m Q_{2n}$ into
$SU(2)$ are distinguished by $\alpha$ and have different discrete indices for a
given irrep ${\bf 2_{k}}$. As for the indices of the one-dimensional irreps we
remark that $\wt \ell({\bf 1_1}) = \wt \ell({\bf 2_0})$ while $\wt \ell({\bf
  1_2}) + \wt \ell({\bf 1_3}) = \wt \ell({\bf 2_n=2_{n\bs\alpha}})$. This
introduces a new parameter $y$ into the definition of the discrete indices\\[-1mm]
\beq
\wt \ell({\bf 1_2}) ~=~y \ , \qquad  \wt \ell({\bf 1_3}) ~=~ \wt \ell({\bf
  2_{n\bs\alpha}}) - y \ .\label{DIQ1}
\eeq

Alternatively, the discrete group $\m Q_{2n}$ could originate from $SU(3)$. In
that case, the embedding would be defined by fixing how the ${\bf 3}$ of
$SU(3)$ breaks into irreps of $\m Q_{2n}$. As we want to embed the complete
group, the decomposition of the~${\bf 3}$ must involve the irrep ${\bf
  2_{\bs\alpha}}$ with $\alpha$ coprime to $2n$. The requirement of the
generators $a$ and $b$ having determinant one then leads to the decomposition\\[-3mm]
$$
{\bf 3} ~\rightarrow ~ {\bf 1_0}+{\bf 2_{\bs\alpha}} \ .
$$
With the methods of Ref.~\cite{Luhn:2008sa} it is straightforward to prove
that the resulting discrete indices are identical to Eqs.~(\ref{DIQ})
and~(\ref{DIQ1}). Thus it is irrelevant whether the discrete symmetry $\m
Q_{2n}$ originates in $SU(2)$ or $SU(3)$. 

Finally, we need to discuss the particles which acquire mass when the
continuous family symmetry is broken to $\m Q_{2n}$ and their effect on the
$\m Q_{2n} - \m Q_{2n} - U(1)_Y$ anomaly. A look at the Kronecker products
reveals that all but one of the bilinear mass terms are obtained from a {\it
  square} 
${\bf r \otimes r}$, so that they do not contribute to the discrete anomaly
condition. The only exception is ${\bf 1_2 \otimes 1_3 =1_0}$ for odd values
of $n$. In order for this not to change the discrete anomaly equation, we must
choose $y$ such that $\wt \ell({\bf 1_2}) = \wt \ell({\bf 1_3})$. Then
Eq.~(\ref{DIQ1}) gets replaced by
\beq
\wt \ell({\bf 1_2}) ~=~ \wt \ell({\bf 1_3})~=~ \frac{n^2}{2} \,+\,\zeta \cdot
2n \qquad(n=\mathrm{odd}) \ ,\label{DIQ1odd}
\eeq
with $\zeta=0,1$, while for even $n$ we still have Eq.~(\ref{DIQ1}) with
arbitrary $y$. With the discrete indices defined in
Eqs.~(\ref{DIQ})-(\ref{DIQ1odd}), a model is discrete anomaly free if 
\beq
\sum_{i=\mathrm{light}} Y_i \cdot \wt \ell_i  ~=~ 0~\mathrm{mod}~4n\ ,\label{Qdac}
\eeq
is satisfied for at least one embedding, i.e. one particular value of
$\alpha$. Let us assume that there exists such an embedding so that
Eq.~(\ref{Qdac}) can be written as
\beqn
\sum_{j=0}^3 c'_j \, \wt \ell({\bf   1_j}) ~+~ \sum_{\kappa=1}^{n-1} c_\kappa
\, \wt \ell({\bf   2_{\bs{\kappa\alpha}}})&=& 0~\mathrm{mod}~4n\ ,\label{Qdac1}
\eeqn
where  the integer coefficients $c'_j$ and $c_\kappa$ are obtained by summing
the hypercharges of all particles living in the corresponding irrep of $\m
Q_{2n}$.  Plugging in the explicit expressions for the discrete indices, we
get
\beq
\!\!\!\left.\begin{array}{l} 
c'_1 \, x
+ \,c'_2 \, y +   c'_3 \, (n^2+x \,i^n-y) 
+ \sum_{\kappa=1}^{n-1} c_\kappa[\kappa^2 + x/2 \, (i^\kappa+i^{-\kappa})]  
\\[3mm]
c'_1 \, \xi\,2n  
+ (c'_2 +   c'_3) \, (n^2/2 +\zeta \, 2n)
+ \sum_{\kappa=1}^{n-1} c_\kappa[\kappa^2 +\xi\,n(i^\kappa+i^{-\kappa})]
 \end{array} \!\!\right\} 
=\, 0~\mathrm{mod}~4n\ ,\nonumber
\eeq
where the first/second line is valid for even/odd $n$. This equation
must hold for all possible values of $x,y\in \mathbb R$ and $\xi,\zeta=0,1$. 
For even $n$ it follows that $c'_2=c'_3=c_n$ as well as $c'_1+c'_3\,i^n +
\sum_{\kappa=1}^{n-1} c_\kappa (i^\kappa+i^{-\kappa})/2=0$.  
For odd $n$ the term $(c'_2+c'_3)\cdot n^2/2$ has to be integer allowing us to
define $2c_n=c'_2+c'_3$ with $c_n\in \mathbb Z$. Thus we can simplify our
equation to 
\beq
\!\!\!\left.\begin{array}{l} 
\sum_{\kappa=1}^{n} c_\kappa \kappa^2 
\\[3mm]
c'_1 \, \xi\,2n  
+ \sum_{\kappa=1}^{n} c_\kappa[\kappa^2 +\xi\,n(i^\kappa+i^{-\kappa})]
 \end{array} \!\!\right\} 
=\, 0~\mathrm{mod}~4n\ .\nonumber
\eeq
Next we multiply everything with $\alpha^2$ and use the fact that $\alpha$ is
necessarily odd so that  $\alpha^2 n = n+(\alpha-1)(\alpha+1)n
=n~\mathrm{mod}~4n$. Therefore, after multiplication, we can remove the factor
$\alpha^2$ in those terms which are proportional to~$n$. Furthermore, observing
that $(i^{\kappa}+i^{-\kappa})= (i^{\alpha\kappa}+i^{-\alpha\kappa})$, we
finally obtain 
\beq
\!\!\!\left.\begin{array}{l} 
\sum_{\kappa=1}^{n} c_\kappa (\alpha\kappa)^2 
\\[3mm]
c'_1 \, \xi\,2n  
+ \sum_{\kappa=1}^{n} c_\kappa[(\alpha\kappa)^2 +\xi\,n(i^{\alpha\kappa}+i^{-\alpha\kappa})]
 \end{array} \!\!\right\} 
=\, 0~\mathrm{mod}~4n\ .\nonumber
\eeq
This is identical to the discrete anomaly equation for the ``standard''
embedding which uses $\alpha=1$. Hence, it is sufficient to evaluate 
Eq.~(\ref{Qdac1}) with $\alpha=1$! If a model is shown to be anomalous for
this embedding, it is automatically anomalous for all other embeddings. Else,
the model is discrete anomaly free. We summarize the discrete indices for the
standard embedding in Table~\ref{tabDIQ}.
\begin{table}[t]
\begin{center}
\begin{tabular}{|c|c|c|}
\hline  &\multicolumn{2}{|c|}{} \\[-2.5mm]
 \raisebox{-1mm}[0mm][0mm]{$\m Q_{2n}$} & \multicolumn{2}{|c|}{$\wt \ell({\bf r})$~~$[\,N_\ell=4n\,]$} \\[-2.5mm]
& \multicolumn{2}{|c|}{}\\ 
\cline{2-3} &  &\\[-2.5mm] 
\raisebox{1mm}[0mm][0mm]{irreps} & $n=\mathrm{even}$ & $n=\mathrm{odd}$  \\[-2.5mm] 
& &\\ \hline && \\[-2.5mm]
${\bf 1_0}$  & ~~~0~~~  & ~~~0~~~    \\[1.5mm]
${\bf 1_1}$  & $x$   & $\xi \cdot 2n$    \\[1.5mm]
${\bf 1_2}$  & $y$  & $n^2/2 + \zeta\cdot 2n$    \\[1.5mm]
${\bf 1_3}$  & $n^2+x\,i^n -y$  & $n^2/2 + \zeta\cdot 2n$    \\[1.5mm]
${\bf 2_k}$  & $k^2+x/2\cdot (i^k+i^{-k})$  & $k^2+\xi\cdot n\,(i^k+i^{-k})$    \\[1.5mm]\hline
\end{tabular}
\end{center}\vspace{-0mm}
\caption{\label{tabDIQ}The discrete indices $\wt \ell$ for the irreps ${\bf
    r}$ of $\m Q_{2n}$ using the standard embedding ($\alpha=1$). The
  discrete anomaly condition has to be satisfied for all possible values of
  $x,y\in \mathbb R$ and $\xi,\zeta=0,1$.} \vspace{2mm}
\end{table}

To conclude our discussion of the  binary dihedral groups, we calculate the
discrete anomaly for some flavor models which rely on the groups $\m Q_{2n}$. 
The models of Refs.~\cite{Frampton:1995gu,Frampton:1996cc} are discrete
anomaly free due to the $SU(5) \times \m Q_{2n}$ grand unified structure.
For the remaining examples, the assignments of the quarks and leptons to the
irreps of $\m Q_{2n}$ are listed in Table~\ref{tabQ}. Other fermions that are
introduced in these models (like for example the Higgs doublets in
supersymmetric models) give no net contribution to the $\m Q_{2n}-\m
Q_{2n}-U(1)_Y$ anomaly. 
Using the indices of Table~\ref{tabDIQ}, one finds that the models of
Refs.~\cite{Chang:1991ri,Frampton:1994xm,Frampton:1994rk,Frampton:1995fta,Babu:2004tn,Frigerio:2004jg,Frigerio:2005pz}
are discrete anomaly free (the assignment in Ref.~\cite{Frampton:1995fta} is
compatible with an $SU(5) \times \m Q_{2n}$ structure, Ref.~\cite{Babu:2004tn}
features a Pati-Salam~$\!\times \,\m Q_6$ compatible assignment), 
while the models proposed in
Refs.~\cite{Chang:1990uga,Kajiyama:2005rk} need additional light particles to
compensate the non-vanishing discrete anomaly. Whether this can be achieved
without contradicting bounds on hypercharged light exotic particles needs an
extensive study of all the possible cases and is therefore left for future
investigations.

\begin{table}[!ht]
{{
\hspace{-10.5mm}\begin{tabular}{|c|c|c|c|c|c|c|c|}
\hline  && &&&&& \\[-3mm]
Group & Refs. & $Q$ & $u^c$ &  $d^c$ & $L$ & $e^c$  & $\sum Y_i \, \wt \ell_i$
\\[-3mm] 
&&&&&&&\\ \hline &&&&&&& \\[-3mm]
$\m Q_4$ & \cite{Chang:1990uga} & $\!{\bf 1_0 , 1_0 ,
  1_0}\!$ & $\!{\bf 1_0 , 1_0 , 1_0}\!$ & $\!{\bf 1_0 , 1_0 , 1_0}\!$ & ${\bf
  1_0 , 2_1}$ & ${\bf  1_0 , 1_1 , 1_3}$ & $2-6y~\mathrm{mod}~8$
\\[-3mm] 
&&&&&&&\\ \hline &&&&&&& \\[-3mm]
$\m Q_4$ & \cite{Frigerio:2004jg,Frigerio:2005pz} & $\!{\bf 1_1 , 1_2 ,
  1_3}\!$ & $\!{\bf 1_1 , 1_2 , 1_3}\!$ & $\!{\bf 1_1 , 1_2 , 1_3}\!$ & ${\bf 1_0 , 2_1}$ & ${\bf  1_0 , 2_1}$ & $0~\mathrm{mod}~8$
\\[-3mm] 
&&&&&&&\\ \hline &&&&&&& \\[-3mm]
$\m Q_6$ & \cite{Chang:1991ri} & $\!{\bf 1_0 , 1_0 , 1_0}\!$ & $\!{\bf 1_0 ,
  1_0 , 1_0}\!$ & $\!{\bf 1_0 , 1_0 , 1_0}\!$ & ${\bf 1_1 , 2_1}$ & ${\bf  1_1
  , 1_2 , 1_3}$ & $0~\mathrm{mod}~12$
\\[-3mm] 
&&&&&&&\\ \hline &&&&&&& \\[-3mm]
$\m Q_6$ & \cite{Frampton:1994xm,Frampton:1994rk} & ${\bf 1_0 , 2_1}$ & $\!{\bf 1_0 , 1_0 , 1_0}\!$ &
${\bf 1_1 , 2_2}$ & ${\bf 1_0 , 2_1}$ & ${\bf  1_1 , 2_2}$ & $0~\mathrm{mod}~12$
\\[-3mm] 
&&&&&&&\\ \hline &&&&&&& \\[-3mm] 
$\m Q_6$ & \cite{Frampton:1994rk} & $\!{\bf 1_0 , 1_0 , 1_0}\!$ & ${\bf 1_0 , 2_1}$ &
${\bf 1_1 , 2_2}$ & $\!{\bf 1_0 , 1_0 , 1_0 }\!$ & ${\bf  1_1 , 2_2}$ & $0~\mathrm{mod}~12$
\\[-3mm] 
&&&&&&&\\ \hline &&&&&&& \\[-3mm] 
$\m Q_6$ & \cite{Babu:2004tn} & ${\bf 1_1 , 2_1}$ & ${\bf 1_3 , 2_2}$ & ${\bf
  1_3 , 2_2}$  & ${\bf 1_1 , 2_1}$ & ${\bf 1_3 , 2_2}$& $0~\mathrm{mod}~12$
\\[-3mm] 
&&&&&&&\\ \hline &&&&&&& \\[-3mm] 
$\m Q_6$ & \cite{Kajiyama:2005rk} & ${\bf 1_1 , 2_1}$ & ${\bf 1_2 , 2_2}$ &
${\bf 1_2 , 2_2}$ & ${\bf 1_0 , 2_2}$ & ${\bf 1_0 , 2_2}$ & $3~\mathrm{mod}~12$ 
\\[-3mm] 
&&&&&&&\\ \hline &&&&&&& \\[-3mm] 
$\m Q_{2n}$ & \cite{Frampton:1995fta} & ${\bf 1_1 , 2_2}$ & ${\bf 1_1 , 2_2}$ &
${\bf 1_0 , 2_1}$ & ${\bf 1_0 , 2_1}$ & ${\bf 1_1 , 2_2}$ & $0~\mathrm{mod}~4n$ \\[1mm]\hline
\end{tabular}}}
\caption{\label{tabQ}The $\m Q_{2n}-\m Q_{2n}-U(1)_Y$ anomaly for various flavor
  models adopting the shown assignments of the quarks and leptons to irreps of
  $\m Q_{2n}$.} 
\end{table}

\section{Conclusion}
In this letter we have derived the discrete indices for the irreps of the
binary tetrahedral group~$\m T'$ as well as the binary dihedral groups~$\m
Q_{2n}$. Despite the ambiguities of embedding the finite symmetries into 
continuous $SU(2)$ and $SU(3)$, it is possible to define discrete indices that
enter the discrete anomaly equations in a non-ambiguous way. Using the results
shown in Tables~\ref{tabDIT} and \ref{tabDIQ}, it is straightforward to
check whether a given flavor model is consistent with a gauge origin of the
applied discrete symmetry.

Having discussed the procedure for the groups $\m T'$ and $\m Q_{2n}$, it
should be clear how to obtain discrete indices for other subgroups of $SU(2)$
like the binary octahedral group and the binary icosahedral group. However, we
are unaware of any example that imposes one of these groups as a family
symmetry. 

\vspace{6.5mm}

{\it Note added:} There has been some confusion as to whether the
results of Ref.~\cite{Araki:2008ek} and Ref.~\cite{Luhn:2008sa} are
compatible. While Ref.~\cite{Araki:2008ek} determines the discrete anomaly by
calculating how the path integral changes under a transformation of the finite
group $\m G$, the approach pursued in Ref.~\cite{Luhn:2008sa} (and also
in this article) relies on embedding $\m G$ into the continuous group
$G_f=SU(3),~SO(3)~\mathrm{or}~SU(2)$. Due to the latter, an anomaly of the
form $G_f - \mathrm{SM} - \mathrm{SM}$ is automatically absent if {\it
 complete} irreps of $G_f$ are considered. On the other hand,
the study in Ref.~\cite{Araki:2008ek} is independent of a specific embedding
and hence does not know of complete or incomplete multiplets of
$G_f$. Therefore no inconsistencies exist between both results.

\section*{Acknowledgments}
I am greatly indebted to Pierre Ramond for countless enlightening
discussions and many helpful comments. This work is supported by the
University of Florida through the Institute for Fundamental Theory.




\end{document}